\documentclass[nofootinbib,superscriptaddress,a4paper,twocolumn, floatfix]{revtex4-2}
\usepackage[english]{babel}
\usepackage[utf8]{inputenc}

\usepackage{amsthm}
\usepackage{amssymb}
\usepackage{mathtools}
\usepackage{physics}
\usepackage{xcolor}
\usepackage{graphicx}
\usepackage[left=23mm,right=13mm,top=35mm,columnsep=15pt]{geometry} 
\usepackage{algpseudocode}
\usepackage{float}
\usepackage{caption}

\usepackage{adjustbox}
\usepackage{placeins}
\usepackage[T1]{fontenc}
\usepackage{lipsum}
\usepackage{csquotes}
\usepackage[pdftex, pdftitle={Article}, pdfauthor={Author}]{hyperref}
\usepackage{xcolor}
\hypersetup{
    colorlinks,
    linkcolor={red!50!black},
    citecolor={blue!50!black},
    urlcolor={blue!80!black}
}
\usepackage{booktabs}
\usepackage{multirow}
\usepackage{listings}
\usepackage{xcolor}
\definecolor{codegreen}{rgb}{0,0.6,0}
\definecolor{codegray}{rgb}{0.5,0.5,0.5}
\definecolor{codepurple}{rgb}{0.58,0,0.82}
\definecolor{tqblue}{HTML}{08293d}
\definecolor{backcolour}{HTML}{fefdf5}

\lstdefinestyle{mystyle}{
    backgroundcolor=\color{backcolour},   
    commentstyle=\color{codegreen},
    keywordstyle=\color{magenta},
    numberstyle=\tiny\color{codegray},
    stringstyle=\color{codepurple},
    basicstyle=\ttfamily\footnotesize\color{tqblue},
    breakatwhitespace=false,         
    breaklines=true,
    postbreak=\mbox{\textcolor{magenta}{$\hookrightarrow$}\space},                 
    captionpos=b,                    
    keepspaces=true,                 
    numbers=left,                    
    numbersep=5pt,                  
    showspaces=false,                
    showstringspaces=false,
    showtabs=false,                  
    tabsize=2
}

\begin{document}

\title{The advent of fully variational quantum eigensolvers\\ using a hybrid multiresolution approach}

\author{Fabian Langkabel}
\email[E-mail:]{fabian.langkabel@algorithmiq.fi}
\affiliation{Algorithmiq Ltd, Kanavakatu 3C, FI-00160 Helsinki, Finland}

\author{Stefan Knecht}
\affiliation{Algorithmiq Ltd, Kanavakatu 3C, FI-00160 Helsinki, Finland}

\author{Jakob~S.~Kottmann}
\email[E-mail:]{jakob.kottmann@uni-a.de}
\affiliation{{Institute for Computer Science, University of Augsburg, Germany }}
\affiliation{{Center for Advanced Analytics and Predictive Sciences, University of Augsburg, Germany }}

\date{\today} % Leave empty to omit a date

\begin{abstract}
In electronic structure theory, variational computing offers a valuable paradigm for the approximation of electronic ground states. However, for historical reasons, this principle is mostly restricted to model-chemistries in pre-defined fixed basis sets. Especially in quantum computation, these model-chemistries are far from an accurate description of the initial electronic Hamiltonian. This work demonstrates a \textit{fully} variational approach to the electronic structure problem by variationally optimizing the orbitals representing the second quantized Hamiltonian alongside a quantum circuit that generates the many-electron wavefunction. To this end, the orbitals are represented within an adaptive multi-wavelet format, guaranteeing numerical precision. We then showcase explicit numerical protocols and highlight the quantum circuit's effects on determining the optimal orbital basis. 
\end{abstract}

\maketitle

%\section{Introduction}

In the long term, the hope for quantum computers is to increase the applicability of (near exact) quantum chemical calculations. A central algorithm here is the quantum phase estimation~\cite{aspuru2005simulated} that can compute accurate eigenenergies of a given second-quantized Hamiltonian. The only remaining source of error w.r.t the original electronic Schrödinger equation is the finite orbital basis, used to construct the second quantized Hamiltonian from the original real-space operator. Execution of such energy estimations is currently anticipated to be computationally expensive~\cite{reiher2017elucidating, vonburg2021quantum, rocca2024reducing} which motivated the development of optimal representations w.r.t to the chosen orbital basis.
The standard approach is to select an active space of orbitals where the quantum computation takes place followed by repeatedly rotating this active space within the fixed space of the basis orbitals.
In order to operate in the optimal orbital basis on a given number of available qubits, essentially two conditions need to be satisfied: 1. the underlying atom-centered Gaussian basis set needs to be sufficient to represent the optimal orbitals, 2. the orbital-optimization algorithm needs to converge. By choosing a large (correlation-consistent) basis set (e.g. d-aug-cc-pCV6Z) the first condition can, in principle, always be met such that no domain expertise is necessary to ensure a sufficient quality of the basis set. The resulting calculations are however far from feasible -- this is due to the increased computational cost of the underlying orbital optimization as well as decreased empirical heuristics to initialize the parameters of the orbital optimization in such large basis sets. The second condition is also challenging, as the selection of active spaces alone is already not trivial~\cite{stein2019autocas, scineautocas} and an ongoing research topic~\cite{stein2016automated, stein2016delicate, sayf17a, mitra2024localizedactivespacemethod}.\\
A prominent alternative to phase estimation are variational algorithms that deal with expectation values of electronic Hamiltonians. In the long term, they are expected to provide useful initial states to increase the success probabilities of direct energy measurement.  Although computationally more tractable, the evaluation of these expectation values and the optimization of the underlying quantum circuits is still considered costly, so that optimal bases are again desirable.\\  
In this work, we will investigate the direct construction of second-quantized electronic Hamiltonians by representing the spatial orbitals in a formally complete basis of multi-wavelets. Starting from an initial guess of spatial functions, the active orbitals are further refined by solving an integral equation defined by the current solution of a variational quantum algorithm. We will start with the technical background in electronic structure and multi-wavelet representations, followed by a brief summary of variational quantum eigensolvers. In Sec.~\ref{sec:methodology} we will provide details of the main algorithms and their implementations and assess their numerical performance by means of molecular examples in Sec.~\ref{chap:numerical}.\\

\section{Technical Background}
This work builds up from a series of prior works that all focus focusing on the determination of numerically exact real-space representations of chemical systems in a black-box fashion. Further below, the broader background is illuminated, while we summarize the developments most important for this work in this paragraph.\\
In Ref.~\cite{kottmann2020direct} Kottmann, Bischoff and Valeev realized the first attempt to determine optimal orbitals in an adaptive real-space basis from scratch, essentially reaching the basis set limit with affordable runtime. Here the correlated many-body method was M{\o{}}ller-Plesset Perturbation Theory (MP2) and the determined orbitals where pair-natural orbitals (PNOs). As a follow-up, Kottmann, Aspuru-Guzik \textit{et.al.} employed the optimal PNOs to construct second-quantized Hamiltonians that can then be tackled with the full toolset of classical quantum chemistry (e.g. coupled-cluster or configuration interaction) as well as variational quantum algorithms, realizing the first hybrid approach that couples adaptive real-space orbital determination with general many-body methods.~\cite{kottmann2020reducing} Here, the notion of a surrogate functional, to determine the optimal orbital-basis was introduced -- in this methodology, the basis error is solely determined by the quality of the surrogate functional. In a further work, the same authors adapted the the variational quantum circuits (many-body part) to the surrogate model (one-body-part) leading to the separable pair-approximation.~\cite{kottmann2022optimized}
In Ref.~\cite{valeev2023direct} Valeev,Harrison,\textit{et. al.} derived the equations to determine optimal orbitals for general many-body methods. In the terminology of the previous works these equation determine an exact surrogate potential depending to the reduced density matrices of an arbitrary many-body method. \\
This work functions as a bridge between the surrogate functional of ~\cite{kottmann2020reducing} and Valeev's functional~\cite{valeev2023direct} where we focus on the application domain of quantum computation and interdependence of the quality of the determined orbital-bases with the used quantum circuits.\\
After initial release of this work, Nibbi, Frediani, Dinvay and Mendl released a preprint~\cite{nibbi2025wavefunction} investigating the combination of Valeev's functional with DMRG solvers for the many-body part.\\
We note that different approaches formulated in real-space~\cite{Haubenwallner_2025, hong2022accurate, genovese2011daubechies}, as well as methods to directly optimize Gaussian based basis-sets (e.g. pioneering works of Tamayo-Mendoza \textit{et. al.}~\cite{tamayo2018automatic} using automatic differentiation, see also \cite{arrazola2021differentiable, zhang2022differentiable} and recent works~\cite{ollitrault2025improving}) have been investigated as well.

In the following we will give an overview over the individual fields of research essential to this work.

\subsection{The Electronic Structure Problem}
The electronic structure problem of quantum chemistry aims to approximate eigenstates of electronic Hamiltonians
\begin{align}
    H = \sum_{k=1}^{N_{\text{e}}} h(r_k) + \frac{1}{2}\sum_{k\neq l}^{N_{\text{e}}} {g(|r_k-r_l|)} \label{eq:Hamiltonian-first-quant}  
\end{align}
with $r_k \in \mathbb{R}^3$ being a spatial coordinate of an electron and $h(r_k)$ the corresponding one-body energy operator $h(r_k) = T(r_k) + V(r_k)$ consisting of the kinetic energy operator $T$ plus an external potential $V$ -- typically a Coulomb potential of point-charges generated by nuclei in a given molecule -- in which the electrons move. The two body operator $g(r_k - r_l) = 1/(\|r_k - r_l \|)$ is the electrostatic repulsion between two electrons at points $r_k$ and $r_l$. In absence of magnetic fields, the Hamiltonian is independent of the electron spin, which is usually introduced ad-hoc to the electronic wavefunction.
The state of an isolated electron can be described as a function $\psi(r, s) = \phi(r)\otimes\ket{s}$, with spatial part $\varphi \in L^2\left(\mathbb{R}^3\right)$\footnote{note that the choice of $L^2$ is somewhat pragmatic here, and the Hilbertspace of $\phi$ can be restricted further as a proper subspace of $L^2$ -- See for example Ref.~\cite{herbst2018thesis} Chapter 2.3.} and spin state $\ket{s} \in \text{span}\left\{\ket{\uparrow}, \downarrow\right\}$. Building up from the one-electron spaces we can construct spaces for many-electron wavefunctions as anti-symmetric tensor products. A more convenient and dominantly used form of the electronic Hamiltonian is in it's second quantized form using the anti-commuting field operators~\cite{jordan1927mehrkorperproblem} $\hat{\psi}(r,s)$ ($\hat{\psi}^\dagger(r,s)$) that annihilate (create) electrons with spin $s$ at spatial coordinate $r$
\begin{align}
    H =& \sum_{s\in \left\{\uparrow \downarrow \right\}}\int \operatorname{d}r\; \hat{\psi}^\dagger(r,s) h(r) \hat{\psi}(r,s) \label{eq:Hamiltonian-second-quant} \\&+ \frac{1}{2}\sum_{s_1,s_2\in \left\{\uparrow \downarrow \right\}} \int \operatorname{d}r_1 \int \operatorname{d}r_2\; \hat{\psi}_1^\dagger \hat{\psi}^\dagger_2 g_{12} \hat{\psi}_1 \hat{\psi}_2 \nonumber
\end{align}
where we used $\hat{\psi}_n \equiv \hat{\psi}(r_n, s_n)$ and $g_{12} = g(|r_1-r_2|)$ for brevity. The operator can be interpreted as probing each spatial point and assigning the corresponding energies if electrons are present. The second quantized form given above in Eq.~\eqref{eq:Hamiltonian-second-quant} is admittedly not ideal for numerical methodologies, such that a more prominent version exploits a discretized form where the field operators are expanded into a set of different operators $a_{k}$ ($a_{k}^\dagger$) that annihilate (create) electrons with spin $s_k$ in the spatial orbital $\phi_k$
\begin{align}
    \hat{\psi}(r,s)&= \sum_k \phi_k(r)\lvert s \rangle a_k \label{eq:orbital-basis0} \\
    \hat{\psi}_k^\dagger(r,s)&= \sum_k \langle s \rvert \phi(r) a_k^\dagger\ .\label{eq:orbital-basis}
\end{align}
The formal assumption in the above expansions is, that the spatial functions $\left\{\phi_k(r)\right\}$ are dense in the three-dimensional Hilbert space $\mathbb{L}^2$. Inserting the expansions  ~\eqref{eq:orbital-basis0} and ~\eqref{eq:orbital-basis}, respectively, in Eq.~\eqref{eq:Hamiltonian-second-quant} leads to the usual form of the second-quantized Hamiltonian 
\begin{align}
   H = \sum_{kl} h_{kl} a^\dagger_k a_k + \frac{1}{2}\sum_{klmn}  g_{mn}^{kl} a^\dagger_k a^\dagger_l a_n a_m\ ,\label{eq:Hamiltonian-second-discretized}
\end{align}
with the tensors $h_{kl} = \langle \phi_k \vert h \vert \phi_l\rangle \langle s_l \vert s_k \rangle  $ and $g_{mn}^{kl} = \langle \phi_k \phi_l \vert g_{12} \vert \phi_m \phi_n\rangle \braket{s_k}{s_m} \braket{s_l}{s_n}$ defined via integrals of the corresponding orbitals and spins with the one- and two-body operators.
Note that a set of suitable basis functions $\phi_k$ has to be introduced here which simplifies the original electronic structure problem so that the Hamiltonians from Eq.~\eqref{eq:Hamiltonian-first-quant} and Eq.~\eqref{eq:Hamiltonian-second-discretized} are not equivalent anymore. The central assumption being, that the chosen set of basis functions represents the chemically relevant wavefunctions well enough that this differences becomes negligible. The most common approach in quantum chemical application is the use of fixed sets of atom-centered basis functions (basis sets).  

\subsection{Multiresolution Analysis}
In the context of this work, we leverage multiwavelets to numerically represent the basis orbitals that define the second-quantized operator in Eq.~\eqref{eq:Hamiltonian-second-discretized}. In the following, the basics of multiresolution analysis (MRA) are captured briefly. The main advantages are that, in contrast to atom-centered orbitals, MRA constructs a complete orthonormal basis of $L^2(D)$~\footnote{The Hilbertspace of quadratically integrable functions on the domain $D$} on a finite simulation domain $D = \left[-L, L\right]^d$ with dimension $d$ that can be truncated in a systematic and individual manner. To illustrate the basics, we will use $d=1$ and $L=1$.
A complete orthonormal basis of $L^2(D)$ is given by the (renormalized and centered at $1/2$) Legendre polynomials $\left\{P_n(x)\right\}_{n\in \mathbb{N}}$. If we truncate this set by restricting the maximal order $n$ to a finite integer $k$, we receive an approximation of $L^2(D)$ spanned by these polynomials $V_0 = \text{span}\left(\left\{P_m(x)\right\}_{m<k}\right)$. By increasing $k$ we can increase accuracy but won't have much control over it. Alternatively, we can keep the initial $k$ fixed and divide the domain $D=[-1,1]$ in the middle. This leads to two new non-overlapping domains $D_1^0 = [-1,0]$ and $D_1^1 = [0,1]$, on which we can define two new sets of polynomials up to order $k$ that span the vectorspaces $V_1^0$ and $V^1_1$ (outside of the corresponding domain, the polynomials will be set to zero). Each of the new sets of polynomials we will derive from the original set by shifting and rescaling: $V_1^0 = \mathrm{span}\left(\left\{\frac{1}{2}P_m(2x-1) \vert_{x\in D_1^0}\right\}_{m<k}\right)$, and correspondingly for $V_1^1$. Our total approximation space is then given by the direct sum of the two spaces representing the left and right parts of the original domain $D$ (note that through the piecewise definition, the corresponding functions remain orthogonal). We can repeat this process recursively in order to locally increase the resolution of the domain $D$. The general approximation space is given by $V_{n}^{l} = \mathrm{span}\left\{2^{n/2} P_k(2^nx - l)\right\}$ that will resolve the subdomain $D_n^l = \left[ l\cdot 2^n , (l+1)\cdot 2^n \right]$ with the rescaled and shifted $k$ polynomials. Key to an automatized truncation of the local resolution are the wavelet spaces $W_n^l$ that span the orthogonal complement between two levels: $V_n^l \oplus W_n^l = V_{n+1}^{2l} \oplus V_{n+1}^{2l+1}$. It suffices to compute the wavelets that span $W_0$ and apply the same shifting and rescaling procedure as for the polynomials before to get the wavelets of $W_n^l$. An arbitrary function $f$ can now be represented by the original $k$ polynomials (the space $V_0$) supplemented with an arbitrary deep refinement of the wavelet-spaces $W_n^l$. A complete representation that is dense in $L^2(D)$ is
\begin{align}
L^2(D) = \overline{V_0 \bigoplus_{n\in \mathbb{N}_0} \left( \bigoplus_{0\leq l \leq 2^n-1}^{} W_n^l \right) }.  
\end{align}
Since each wavelet space has two children we can represent this construction as a binary tree. A function $f$ can now be represented with arbitrary accuracy by starting at the root (or a full-tree arbitrary level $n$ to avoid undersampling) and recursive refinement. Once the norm of wavelet coefficients of a specific level $W_n^l$ falls below a given accuracy threshold the refinement can be stopped.\\
\noindent 
Higher dimensional representations can be constructed via specific tensor products between the low-dimensional spaces. A 2-dimensional MRA can for example be constructed as $V_0^{(d=2)} = V_0^{d=1} \otimes V_0^{d=1} = \left\{ P_{m_1}(x)P_{m_2}(y) \right\}_{m_1,m_2<k}$. The shifting and rescaling procedure stays the same, just that we now have 4 children of each parent space splitting the two-dimensional domain into 4 quadrants. Correspondingly there are also 4 wavelet spaces, constructed from the direct tensor product of the $W_0$ and the three possible combinations of $V_0$ and $W_0$. Operators follow essentially the same construction pattern and the details are omitted here. 

\noindent 
An MRA is a proper basis of $L^2$ and is therefore, in the context of quantum chemistry, often referred to as a \textit{basis-set-free} representation. Software packages like \textsc{madness}~\cite{harrison2016madness} or \textsc{mrchem}~\cite{wind2022mrchem, bjorgve2024vampyr} implement high-level access to MRA technology in an intuitive way that is sometimes referred to as a \textit{basis-free} formulation, as the user can choose to ignore the details of the underlying numerical representation entirely. Consider for example two analytically known functions and their product
\begin{align}
    f(x) = e^{-|x|},\;\; g(x) = x^2,\;\; p(x) = g(x)\cdot f(x)
\end{align}
a \textit{basis-free} implementation would take the form
\begin{lstlisting}
def functor_f(x) = exp(-abs(x))
def functor_g(x) = x**2
f = project(functor_f)
g = project(functor_g)
p = f*g
\end{lstlisting}
The so constructed MRA representation of $f,g,p$ is carried out fully automatized in the background and in-praxis guarantees accurate representation up to a given threshold.

\subsection{Multiresolution Quantum Chemistry}
 A general procedure, inherent to most MRA-based methods in quantum chemistry is
\begin{enumerate}
    \item Choose an initial guess of orbitals $\left\{\phi_i\right\}$ and represent each of them through an MRA tree. 
    \item Construct the defining equations for each orbital in the form 
    \begin{align}
        \phi_k(r) = \int G(r-r') V_k\left[\left\{\phi_i \right\}\right](r') \operatorname{d}r'\ ,
        \label{eq:orbupdate}
    \end{align}
     where the potential $V$ usually depends on all orbitals.
    \item Solve Eq.~\eqref{eq:orbupdate}\ by iteratively computing the potentials $V_k$, updating the orbitals according to the defining equation and orthonormalizing the updated orbitals.   
\end{enumerate}
% using this to define notation below
For example, in the case of Hartree-Fock (HF) the potential $V$\ takes the form~\cite{harrison2004multiresolution, yanai2004exchange, jensen2023kinetic}
\begin{align}
    V_{{\text{HF}}_i}(r) = 2\left(\sum_k g_k^k(r)\right) \phi_i(r) - \sum_{k} g_{i}^k(r) \phi_k(r) 
\end{align}
with the abbreviation
\begin{align}
    g_k^l(r) = \left(\int g_{12}(r-r') \phi_k(r')\phi_l(r')  \operatorname{d}r'\right).
\end{align}

In passing we note that in quantum chemistry HF is usually only the first step of more complex calculations due to a poor description of the electron correlation. In order to describe the missing correlation, these methods include additional, initially unoccupied orbitals, the so-called virtual orbitals, in the many-body wavefunction. While these virtual orbitals are obtained as a by-product of the Roothaan-Hall procedure (HF in a fixed basis), this is not the case with MRA representations that typically tackle the HF equations directly by iteratively solving the fixed-point equation~\eqref{eq:bsh_update}. The challenge for post-HF methodology is to generate a set of virtual orbitals that can describe the missing correlation as accurately as possible with as few orbitals as possible. In the early days of MRA quantum chemistry this was often circumvented by solving the corresponding methods -- such as MP2~\cite{bischoff2012computing,  bischoff2013computing} and CC2~\cite{ kottmann2018coupledGS, kottmann2017coupledES, kottmann2018thesis} -- directly by representing the occurring electron-pair functions with six-dimensional MRA trees. The first demonstration of a post-HF with adaptively refined one-body functions was a many-body perturbation method (PNO-MP2-F12)~\cite{kottmann2020direct}. For this purpose, pair-natural orbitals (PNOs) were generated as virtual orbitals, which represent the optimal solution to the problem (PNOs used in this work, will always refer to the PNOs of Ref.~\cite{kottmann2022optimized} obtained without explicit correlation and using the ``diagonal'' approximation). 

More recently, going beyond single-reference ans{\"a}ze, Valeev and co-workers developed a procedure for describing arbitrary MRA-based many-body wavefunctions together with a protocol to determine optimal virtual orbitals.~\cite{valeev2023direct}. In the latter work, they used Heat-Bath CI (HCI) as an exemplary post-HF method and started with virtual orbitals based on finite (Gaussian-type) basis sets.
Apart from multiwavelets, approaches based on Daubechies Wavelets~\cite{hong2022accurate} and optimised Gaussian basis sets via gradient based approaches~\cite{tamayo2018automatic, arrazola2021differentiable, barison2020quantum} have been investigated as alternatives to fixed Gaussian basis sets. In contrast to the present work they focus on the optimization for a mean-field HF wavefunction while approaches to general electronic structure methodology which could not been demonstrated so far. For example, in Ref.~\cite{barison2020quantum} the orbitals from a Gaussian basis set were contracted through intrinsic contraction. Moreover, in contrast to this work, where multiwavelets in a discontinuous Galerkin fashion are used to represent orbitals that form the second quantized Hamiltonian, Ref.~\cite{McClean2020discontinuous} investigated a direct construction in the numerical basis. \\

Apart from traditional ``wavefunction'' based methodologies, the MRA representation has been applied to density functional theory~\cite{yanai2004exchange, jensen2017elephant, hurtado2024benchmarking, pitteloud2023quantifying}, orbital effective potentials~\cite{stueckrath2021reduction}, range-separation~\cite{yanai2005multiresolution, poirier2024range}, magnetic and optical properties~\cite{jensen2016magnetic, bischoff2020structure, brakestad2020static}, solvation models~\cite{cammi2002second, gerez2023cavity} and even relativistic approaches~\cite{anderson2019dirac, anderson2020real, brakestad2024scalar, tantardini2024full, remigio2017four}. Although seemingly unrelated at first glance, these developments, and their open-source implementations within the two main packages \textsc{madness}~\cite{harrison2016madness, anderson2019derivatives} and \textsc{mrchem}/\textsc{vampyr}~\cite{wind2022mrchem, bjorgve2024vampyr} are crucial to this work as well as open up for possible future extensions of our present approach beyond pure ground-state energy estimations.\\

\noindent
To summarize, our algorithm, detailed in Section \ref{sec:methodology}, exploits ideas from previous works in multiresolution quantum chemistry. Starting from HF MRA orbitals~\cite{harrison2004multiresolution, bischoff2014regularizing}, PNOs~\cite{kottmann2020direct} are generated and optimized via Eq.~\eqref{eq:bsh_update} to serve as initial functions for the methodology in Sec.~\ref{sec:methodology}. Subsequently, the $h_{kl}$ and $g_{mn}^{kl}$ tensors are obtained directly by integration on the MRA representation and the resulting Hamiltonian is used to calculate many-body wavefunctions and reduced density matrices (RDMs) via a second-quantized variational procedure.~\cite{kottmann2020reducing}. To optimize many-body wavefunctions we employ (i) a variational quantum eigensolver (see Section~\ref{sec:vqe}) and (ii) a standard full CI solver. The calculated RDMs $\eta_k^l$ and $\eta_{kl}^{mn}$ are then used to refine the MRA orbitals similar to Ref.~\cite{valeev2023direct} but using our own implementation.

\subsection{Variational Quantum Eigensolvers}\label{sec:vqe}
The basic idea of variational quantum computing can be summarized into a three-step procedure: 1. represent the electronic Hamiltonian (Eq.~\eqref{eq:Hamiltonian-second-discretized}) of $N$ spin orbitals as an operator, commonly referred to as qubit Hamiltonian, on $N$ qubits. 2. Construct a parametrized quantum circuit, execute it on a quantum processor and measure the expectation value of the qubit Hamiltonian. 3. Update the parameters of the circuit to lower the energy expectation value and repeat until convergence. Fortunately, the gradients of expectation values can be measured as well~\cite{schuld2019evaluating, kottmann2021feasible}, so that standard classical optimization procedures can be employed. After the optimal parameters of the circuit are determined, it can be used to prepare the electronic ground state approximation $\Psi$ and general observables can be measured. In this work, the observables of interest are the RDMs, that can be written as expectation values w.r.t the quantum state $\Psi$
\begin{align}
    \gamma_k^l = \langle a^\dagger_l a_k \rangle_\Psi,\;\;\gamma_{kl}^{mn} = \langle a^\dagger_m a^{\dagger}_n a_k a_l \rangle_\Psi,\label{eq:rdms}
\end{align}
where we will explicitly require spin-summed reduced density matrices that only depend on the indices of the spatial orbitals
\begin{align}
\eta^l_k = \sum_{\sigma \in \left\{ \uparrow, \downarrow \right\}} \gamma^{l\sigma}_{k\sigma},\;\;\eta^{mn}_{kl} = \sum_{\sigma,\tau \in \left\{ \uparrow, \downarrow \right\}} \gamma^{m\sigma n\tau}_{k\sigma l\tau}.\label{eq:rdms-spin-summed}
\end{align}

\noindent
The first step (encoding the second-quantized Hamitonian) is to represent it in the standard Pauli form
\begin{align}
    H = \sum_k c_k P_k
\end{align}
where $P_k$ are Pauli-strings (tensor products of the Pauli matrices $X,Y,Z$).
The most common encoding is Jordan-Wigner, where we the second quantized operators are encoded quite naturally into the qubit raising and lowering operators $\sigma^\pm = \frac{1}{2}\left( X \pm iY \right)$ and a string of $Z$ operators to ensure anti-commutativity
\begin{align}
    a^\dagger_k = \left(\prod_{l<k} Z_l \right) \sigma^-_k,\;\; a_k = \left(\prod_{l<k} Z_l \right) \sigma^+_k.  
\end{align}
In the same way, we can then express the RDMs of Eq.~\eqref{eq:rdms} in the standard Pauli form. \\

\noindent
The design of the quantum circuit is currently an active problem in research, and several methods have been proposed -- see Ref.~\cite{anand2022quantum, tilly2020computing} for recent reviews. A significant fraction takes inspiration from unitary coupled-cluster, that provides unitary building blocks in the form of $n$-electron excitations. We can construct them with a list of tuples $[(k_0,l_0),(k_1,l_1),\dots]$ that encode electronic exchanges between spin-orbitals $(k_j,l_j)$ and result in the generator
\begin{align}
    G_n = i\left(\prod_j \left(a^\dagger_{k_j} a_{l_j}\right) - h.c.\right)
\end{align}
that itself generates the parametrized unitary
\begin{align}
    U(\theta) = e^{-i\frac{\theta}{2} G}.\label{eq:ccU}
\end{align}
The anti-hermitian as $i(x-x^\dagger)$ ensures that the resulting unitary only creates real superpositions.
The generators $G_n$ can be decomposed into the standard Pauli form in the same way as it was done for the electronic Hamiltonian -- in this case, the decomposition will result in a set of commuting Pauli strings~\cite{romero2018strategies}, so that the compilation of a quantum circuit is straightforward (see for example~\cite{anand2022quantum} or ~\cite{yordanov2020efficient} for a direct approach).\\
\noindent
Pioneering works proposed adaptions of the conventional UCCSD~\cite{peruzzo2014variational,mcclean2016theory, romero2018strategies, grimsley2019trotterized} into the individual building blocks in Eq.~\eqref{eq:ccU} via Trotter decompositions. Over the following years ``bottom-up'' strategies that assemble the UCC building blocks directly -- e.g. through adaptive~\cite{grimsley2019adaptive, ryabinkin2018qubit} or heuristic~\cite{Izmaylov2020order, anselmetti2021local, burton2022exact, burton2024accurate, Ghasempouri2023modular} strategies -- have been demonstrated to be more effective in the cost to benefit ratio.\\

\noindent
In this work, we will employ mostly the separable pair approximation (SPA)~\cite{kottmann2022optimized} and some extensions of it, as the range of applicability of the approximation is understood and the method has shown to produce consistent results in several applications~\cite{gratsea2024comparing, schleich2021improving, schleich2023partitioning, weber2022toward}. This allows us to create wavefunctions with controlled quality in their approximations and study the effects on the basis refinement.  

\section{Fully-Variational Quantum Chemistry}\label{sec:methodology} 

The overall goal of the electronic structure ground state problem given in Eq.~\eqref{eq:Hamiltonian-second-quant}\ can be rephrased to \textit{fully} minimize the many-body energy
\begin{equation}
E\left[\Psi,\left\{\phi_k\right\}\right] = \sum_{kl}\eta_k^l\hat{h}_k^l + \frac{1}{2} \sum_{klmn}\eta_{kl}^{mn}g_{kl}^{mn},
\end{equation}
here given as a functional of a second-quantized many-body wavefunction $\Psi$ and a spatial orbital basis $\left\{ \phi_k \right\}$.
The one- and two-body RDMs $\eta_k^l$ and $\eta_{kl}^{mn}$ (cf.~Eq.~\eqref{eq:rdms}), respectively, are obtained from $\Psi$, while the one- and two-body coefficients $h^l_k$, $g_{kl}^{mn}$ are dependent on $\left\{\phi_k\right\}$\ as can be seen in Eq.~\eqref{eq:orbital-basis}. 
By invoking the variational principle
\begin{align}
    E_0 \leq \min_{\Psi, \left\{\phi_k\right\}} E\label{eq:energy-functional}
\end{align}
the latter will yield the true ground state energy $E_0$ with respect to the original first-quantized Hamiltonian given in Eq.~\eqref{eq:Hamiltonian-first-quant}.
To achieve this goal, we set out from a sequential minimization of a 
parametrized many-body wavefunction $\Psi(\boldsymbol{\theta}$) for a fixed set of orbitals in the form of a VQE run, followed by a variation of the orbital basis for a fixed set of RDMs. The resulting procedure is then repeated until convergence. Whereas the first step is outlined already in Section \ref{sec:vqe}, the orbital variation step deserves further discussion.\\

While the many-body wavefunction is parametrized in a discretized form through parametrized quantum gates, the variation of the orbitals is performed in a continuous fashion. The Lagrangian
\begin{equation}
\mathcal{L} = E\left[\Psi,\left\{\phi_k\right\}\right] -\sum_{kl} \epsilon_{kl} \left( \langle \phi_k | \phi_l \rangle - \delta_{kl} \right)
\end{equation}
results directly from the energy functional in Eq.~\eqref{eq:energy-functional} but also takes the orthonormalization of the spatial orbital basis $\left\{ \phi_k \right\}$ explicitly into account through the last term.
Stationary conditions can now be obtained by requiring the functional derivative to vanish
\begin{align}
    \frac{\delta}{\delta \phi_i} \mathcal{L} = 0.
\end{align}
Taking the functional derivative of the first term in the Lagrangian will reveal the Laplacian hidden in the one-body integrals
\begin{align}
    \frac{\delta}{\delta \phi_i } \sum_{kl} h_k^l \eta_l^k &= \sum_{kl} \frac{\delta}{\delta \phi_i (r)}
    \int \phi_l(r') h(r') \phi_k(r') \operatorname{d}r' \eta_l^k\nonumber\\
    &= \sum_{kl} 2 h(r) \phi_k(r) \delta_{il} \eta_l^k \nonumber\\
    &= 2\sum_k \left(-\frac{1}{2}\nabla^2\phi_k(r) +  V(r)\phi_k(r) \right) \eta_{i}^{k}.
\end{align}
The stationary condition for the whole Lagrangian is then given by the differential equation 
\begin{align}
    2\sum_k \left(-\frac{1}{2}\nabla^2\phi_k(r) +  V(r)\phi_k(r) \right) \eta_{i}^{k} + \frac{\delta}{\delta \phi_i} \mathcal{V} = 0
\end{align}
where $\mathcal{V}$ comprises the second and third term in the energy functional of Eq.~\eqref{eq:energy-functional} given above. 
Following Valeev and co-workers~\cite{valeev2023direct} we can transform the orbitals into natural orbitals by diagonalizing the reduced one-body density matrix $\eta^k_i$  which will eliminate all Laplacian terms with $k\neq i$ in the sum above. This results in the defining equations for the stationary orbitals
\begin{equation}
\phi_i = -2\hat{G}_{-\frac{\epsilon_i^i}{\eta_i^i}} \left(\hat{V}\phi_i - \frac{1}{\eta_i^i} \sum_{k \neq i} \epsilon_k^i \phi_k + \frac{1}{\eta_i^i} \sum_{lnk} g_l^n \eta_{kl}^{in} \phi_k \right),\label{eq:bsh_update}
\end{equation}
where $\hat{G}_\mu=\left(\nabla^2 -\mu\right)^{-1}$ is the bound-state Helmholtz Green's operator and the Lagrange multipliers $\epsilon_a^i$ are given as
\begin{equation}
\epsilon_a^i = \eta_i^i h_a^i + \sum_{lnk} \eta_{kl}^{in} g_{al}^{kn}.
\end{equation}
In a general form, Eq.~\eqref{eq:bsh_update} can be written as
\begin{align}
    \phi_i = -2 \hat{G}_{\mu_i} \mathcal{V}_i\left[\left\{\phi_k\right\}\right]
\end{align}
with a potential operator $\mathcal{V}_i\left[\left\{\phi_k\right\}\right]$ for each orbitals $\phi_i$ that is also functional of all orbitals.
Using the terminology of Ref.~\cite{kottmann2020reducing} this is the surrogate potential that determines the basis of the electronic system. 
Valeevs potential~\eqref{eq:bsh_update} denotes an exact potential, while the PNO-based potential of Ref.~\cite{kottmann2020direct} are an approximation. 
In the same way, a mean-field potential could be used, but this typically results in unbounded virtual orbitals -- \textit{i.e.} for most molecular systems we would only get $N_e/2$ bound solutions.

\begin{figure}
\centering
\includegraphics[width=0.48\textwidth]{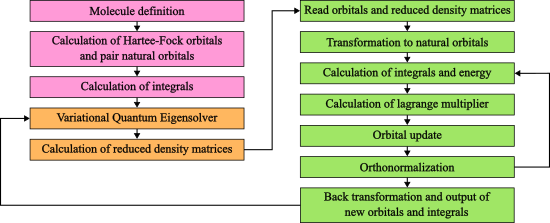}
\captionof{figure}{Flowchart of the proposed multiresolution quantum framework. Based on initial HF orbitals and pair natural orbitals as well as their integrals with the molecular Hamiltonian (pink), energy and RDMs are determined using a VQE ansatz (orange) alternating with an orbital refinement step (green).}\label{fig:flowchart}
\end{figure}

Since the orbital update of each orbital also depends on all other orbitals, the update is performed iteratively until all orbitals converge. For faster convergence, more efficient solvers can be used instead of the fixed point update shown in Eq.~\eqref{eq:bsh_update} -- historically the KAIN~\cite{harrison2004krylov} approximation has proven useful here. Since the update of all orbitals is performed simultaneously without updating the integrals within an iteration step, the orthonormality of the orbitals is no longer given after an iteration step, which makes a reorthonormalization after each iteration step necessary. To this end, we make use of the symmetric (``L{\"o}wdin") orthogonalization following Ref.~\citenum{kottmann2022optimized}. After convergence of all orbitals, the new $h_{kl}$ and $g_{mn}^{kl}$ tensors are used to update the Hamiltonian of the quantum part for optimization of the many-body wavefunction $\Psi(\boldsymbol{\theta}$). The procedure of adapting the many-body wavefunction and orbital refinement is then repeated sequentially until the energy converges to a given threshold. We illustrate our proposed multiresolution quantum computing framework in the flowchart ~\ref{fig:flowchart}. 

In passing we note that our proposed workflow bears a close resemblance to the Complete Active Space Self Consistent Field (CASSCF) approach commonly used in quantum chemistry, in which both the many-body wavefunction and the orbitals are optimized in a similar fashion. In contrast to CASSCF and the use of (Gaussian-type atom-centered) basis sets, our orbital refinement step  
is not limited to the space spanned by the basis functions, which means that unmatched results can be obtained with the same number of orbitals. To this end, we are discriminating between orbital optimization -- rotating the current set of active orbitals within the given orbital basis, and orbital refinement where the orbital basis gets refined with the methodology described above. 
As with orbital optimization, orbital refinement with MRA orbitals can be combined with any other quantum chemistry method which allows the calculation of RDMs, for example, coupled cluster methods, configuration interaction or tensor-network approaches such as DMRG.

\section{Numerical Demonstrations}\label{chap:numerical}

The following subsections summarize numerical demonstrations of our proposed multiresolution quantum  framework described in Section~\ref{sec:methodology} on selected test systems.
We commence with one of the largest possible systems for the orbital refinement implementation presented in this work, cyclohexane with 36 electrons active electrons in 36 spatial (72 spin-) orbitals. Subsequently, we highlight the fundamental features of orbital refinement compared to the previous PNO approximation on simple (effective) two-electron model systems (H$_2$, LiH) before  continuing with chemically more interesting systems (BeH$_2$ and H$_4$). In the latter, we will in particular highlight the influence of an approximate VQE Ansatz (leading to a qubit wavefunction with varying quality across the molecular instances) on the basis refinement. 
\\

\begin{figure}
\centering
    \includegraphics[width=0.5\textwidth]{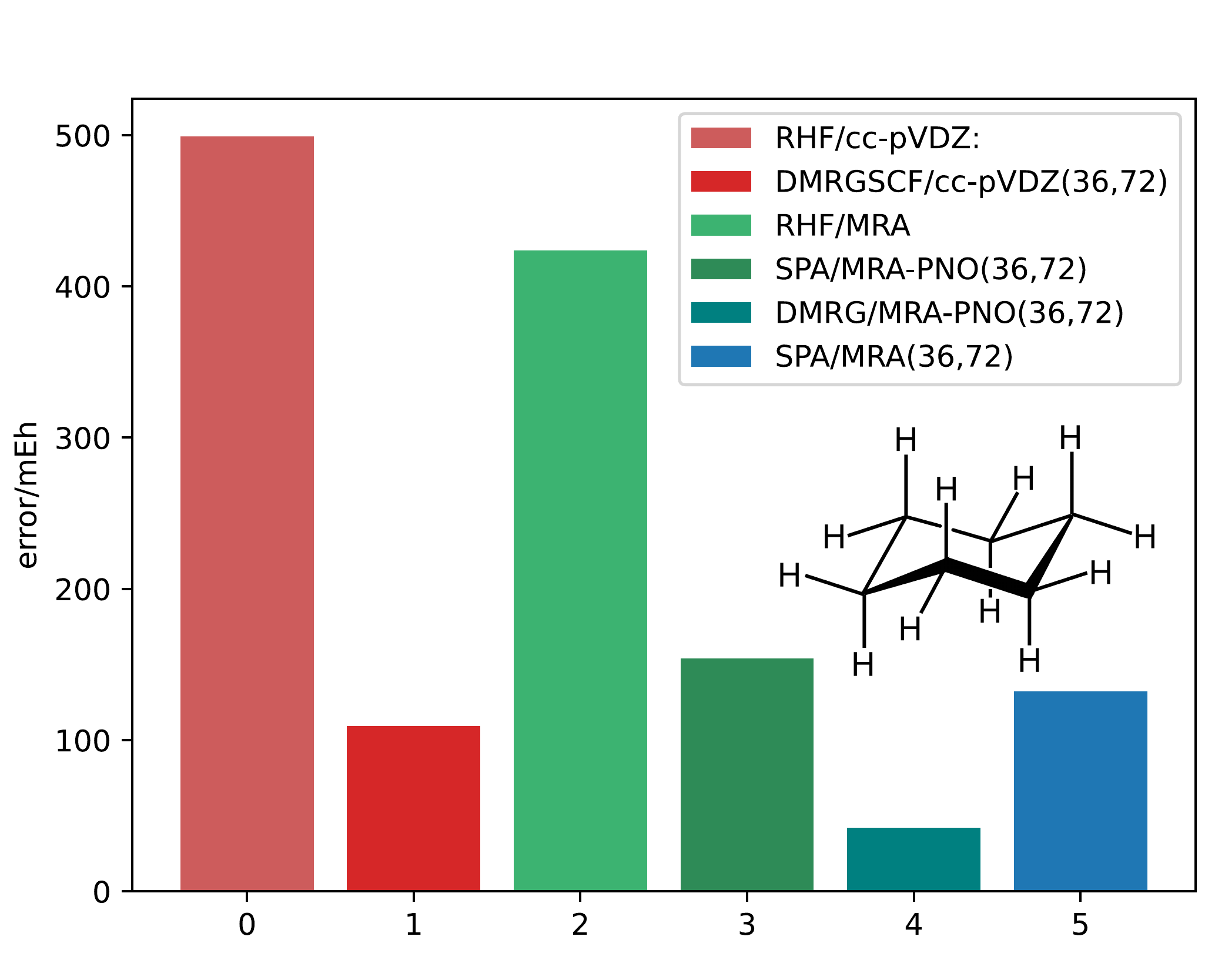}
    \caption{\textbf{Applicability}: Full calculation of the Cyclohexane molecule in an active space of 36 electrons and 36 spatial (72 spin) orbitals with different orbitals and many-body methods. This serves as a stand-in of the system sizes still treatable with the method. We show the energy difference (atomic units) with respect to DMRG[BD=1000]/MRA(36,72). In this example, MRA-PNOs are computationally cheapest while still comparable in numerical precision.}\label{fig:cyclo}
\end{figure}

\noindent
In Tab.~\ref{tab:notation} we summarize the notation used within this work. It follows the standard pattern: WFN/BASIS($N_e$, $N_q)$, where WFN refers to the method employed to represent the second quantized many-electron wavefunction (or quantum circuit) and BASIS refers to the method employed to determine the one-electron basis functions (spatial orbitals) that define the second quantized Hamiltonian~\eqref{eq:Hamiltonian-second-discretized}. The numbers $N_e$ and $N_q$ refer to the number of active electrons and qubits, respectively (the latter being equal to the number of spin-orbitals) for our multi-configurational wavefunction \textit{ansatz}.\\

\noindent For standard Gaussian basis sets, BASIS is just the identifier of the corresponding basis set. For example, FCI/cc-pVDZ($N_e$,$N_q$) denotes a full CI calculation, that is, an exact diagonalization in the $N_e$-electron singlet subspace,  and $N_q$ will refer to two times the number of basis orbitals in the standard basis set cc-pVDZ (corresponding to the number of qubits in most encodings).
By contrast, for MRA represented orbitals we use the general notation: MRA[wfn, opt, it] where ``wfn''denotes the many-electron wavefunction used in the basis refinement, ``opt'' denotes the number of orbitals that are refined, and ``it'' denotes the number of macro iterations. For better readability, we list in Table~\ref{tab:notation} all  abbreviations of this general format.\\

\noindent
In appendix~\ref{sec:comp-details} we provide all computational details for the numerical studies in this work along with references to the employed software packages.

\begin{table}
\caption{\textbf{Notation:}  Methods are denoted as WFN/BASIS($N_e$,$N_s$) where WFN denotes the method to describe the many-body wavefunction (e.g. SPA+GSD, FCI, CASSCF) while BASIS denotes the various orbital bases listed in the table. See Sec.~\ref{chap:numerical} for some examples.}\label{tab:notation}
\begin{tabular}{ l l}
\toprule
\midrule
Basis identifiers & Details \\
\midrule
PNO & Short for MRA-PNO as in Ref.~\cite{kottmann2022optimized} \\
MRA & orbitals refined as in Sec.~\ref{sec:methodology}\\
MRA[it=$n$] & Results after $n$ macro-iterations\\
MRA[opt=$n$] & Only the first $n$ orbitals are refined\\
MRA[wfn=$X$] & Different wavefunction used for refinement\\
(aug-)cc-pVXZ & Reference Gaussian basis sets~\cite{dunning1989gaussian}\\
\midrule
\bottomrule
\end{tabular}
\end{table}

\begin{figure*}
\centering
\begin{tabular}{lr}
\includegraphics[width=0.45\textwidth]{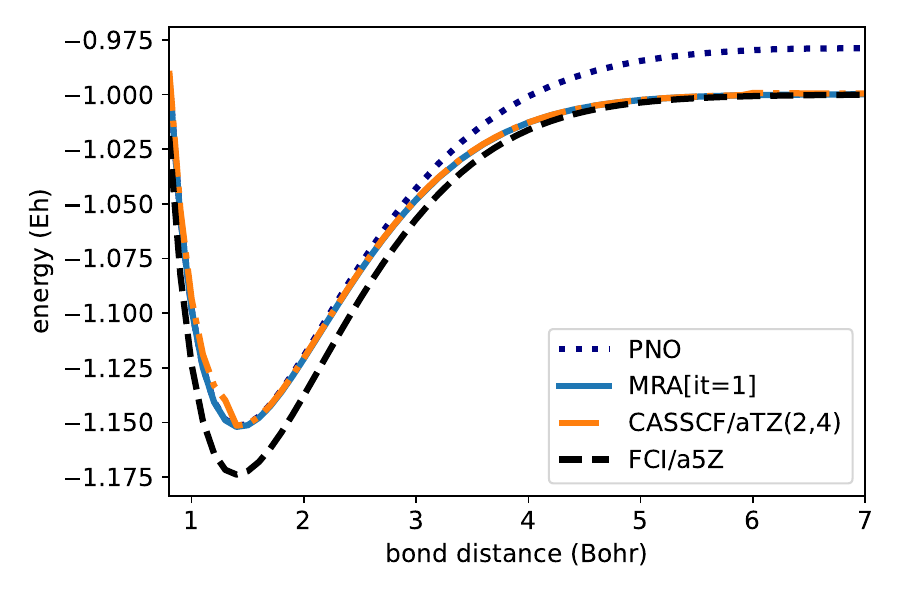}&
\includegraphics[width=0.45\textwidth]{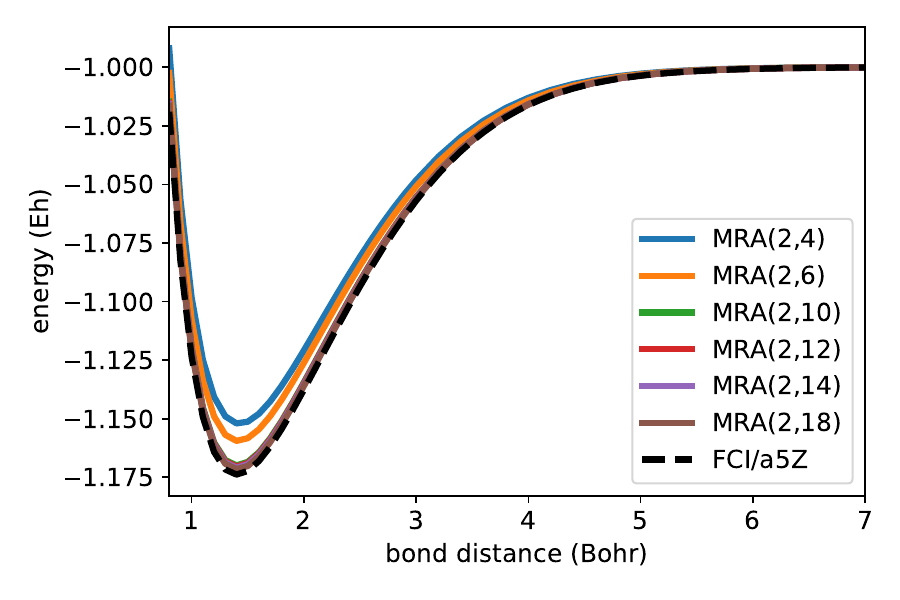}\\
\includegraphics[width=0.43\textwidth]{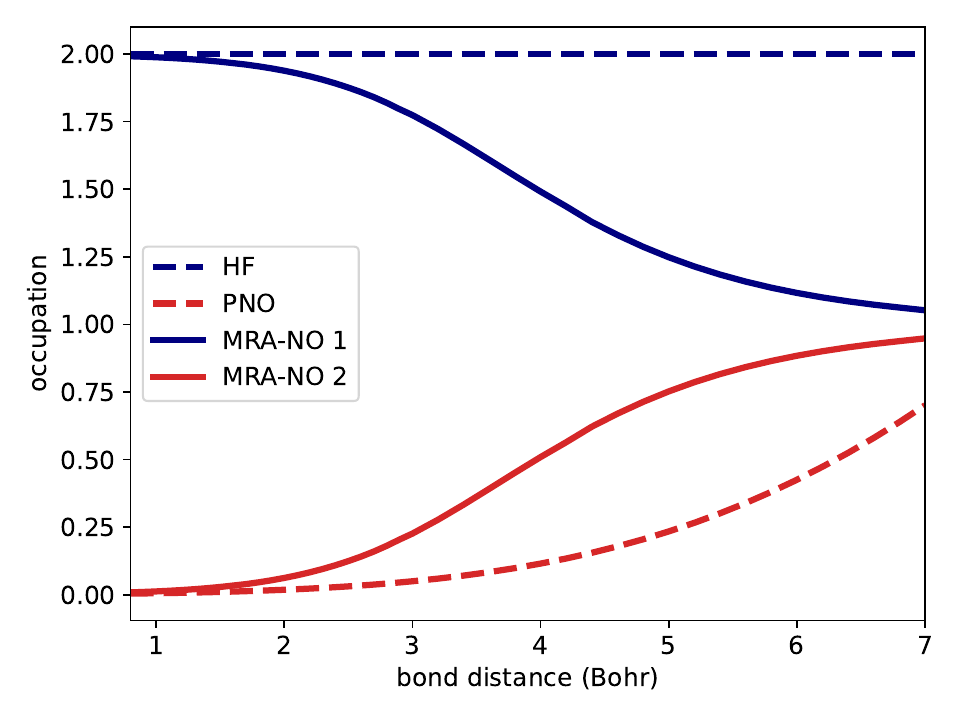}&
\includegraphics[width=0.43\textwidth]{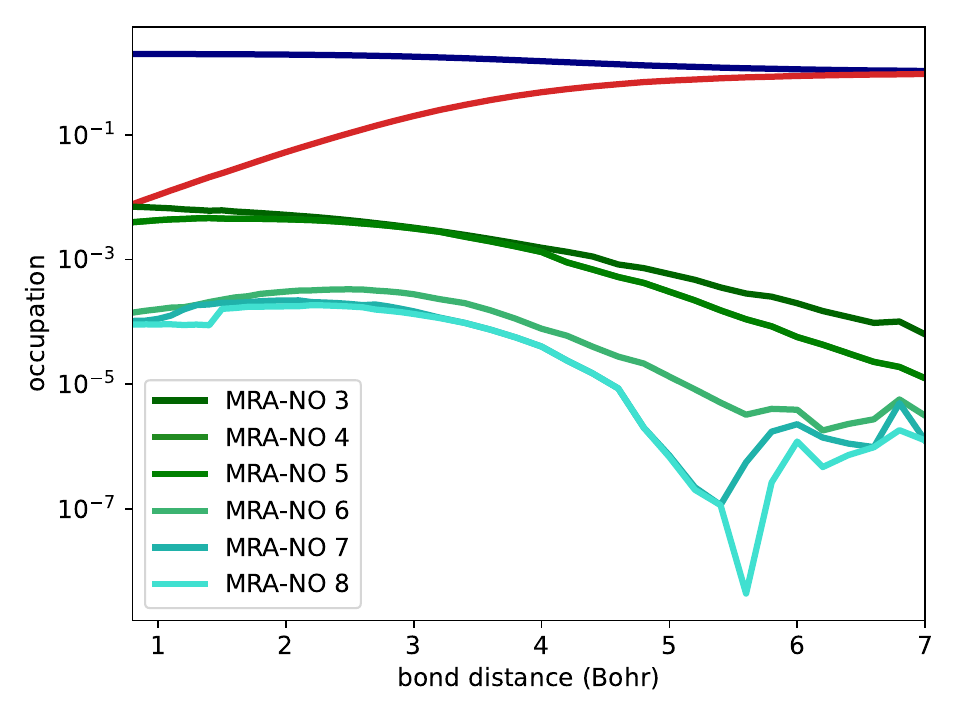}\\
\end{tabular}
\captionof{figure}{\textbf{Illustration on the H$_2$ molecule:} Comparison of the H$_2$ dissociation calculated with SPA+GSD(2,4) and adaptive MRA orbitals after several optimization steps starting from pair natural orbitals as an initial guess. Top row: total energies. Bottom row: occupation numbers.}\label{fig:h2(2,4)}
\end{figure*}

\subsection{Cyclohexane}
We used the cyclohexane molecule as a demonstration of a molecule that is still feasible to compute with exact basis refinement and will give an exemplatory illustration of the cost and perspective gains of orbital refinement. The results are shown in Fig.~\ref{fig:cyclo} To the best of our knowledge, this is one of the largest simulations for VQE optimizations as well as exact orbital determination.\\
When compared to the cheaper MRA-PNOs, the total error is between 20-40 mEH depending on the used many-body state: in this case either a SPA quantum cirucit or a DMRG calculation with large bond dimension. The determination of the MRA-PNOs takes around 15 minutes on a standard compute node, while the full refinement coupled to a DMRG calculation takes around 1-2 days. These timings should not be taken as accurate benchmarks but rather as illustrations of the necessary cost to reach the last 20-40 mEh in the energy. In this case, a full basis refinement with quantum circuits on SPA level alone is most likely not an attractive scenario, as the method-error of the SPA circuits alone is already around 100 mEh which is significantly higher as the corresponding basis error of around 20 mEh.\\

To put the numbers into perspective: An orbital-optimized active space simulation within the cc-pVDZ basis has already a significantly larger basis error of around 100 mEh (75 of them already at the mean-field level). In our case the calculation took around 15 hours and was significantly longer than the MRA-PNOs that took around 15 minutes (with around the same time for the subsequent many-body simulation). 

\subsection{H$_2$ and LiH}
To study the influence of orbital optimization on the energy and the many-body wavefunction, we calculated the dissociation curves of H$_2$ and LiH with different active spaces. For both molecules, SPA circuits can describe the (effective) two-electron systems exactly if the orbitals are in the correct linear combination.~\cite{kottmann2022molecular} We use a slightly expanded approach in the form of SPA+GSD (SPA plus a layer of generalized pair-excitations that are missing in the leading SPA excitations followed by a layer of generalized singles excitations) to aid the orbital-optimizer in finding the correct linear combinations.\\
To verify whether and how fast the orbitals and the multi-particle wavefunction expressed by reduced density matrices converge with alternating optimization, the dissociation curve of H$_2$ with 4 spin-orbitals was considered as the simplest possible test case with results displayed in Fig.~\ref{fig:h2(2,4)}.
In the case of the dissociated H$_2$, the system is resolved exactly with two spatial orbitals given by the analytic solution of the isolated hydrogen atom. 
The spatial orbital basis determined via the MP2-PNO model is not able to represent this solution due to the neglect of important coupling terms in the MP2 model. Hence, a refinement of the PNO basis via Eq.~\eqref{eq:bsh_update} is able to correct this deficiency in a single iteration. 
By comparison, at shorter bond distances, the energy of the refined basis is of similar quality as the CASSCF/aug-cc-pVTZ(2,4) reference which indicates that this is the best possible representation restricted to two spatial orbitals only and the resulting error is solely due to this restriction. We confirm the latter hypothesis by increasing the number of spatial orbitals in the active space from two to 9. As illustrated in Fig.~\ref{fig:h2(2,4)} this modest increase results in a smooth convergence towards the full FCI/aug-cc-pVTZ(2,92) energy (obtained with 46 orbitals). In passing we note that the remaining total error is lower than 3 millihartree compared to a aug-cc-pV5Z reference indicating that we can essentially reach the basis set limit for this system. \\

The convergence behavior in the case of the H$_2$ molecule  can be explained as follows: In a representation with two spatial natural orbitals $\phi_0$ and $\phi_1$, the full first-quantized wavefunction of H$_2$ can be expressed as a linear combination of two anti-symmetric pair functions like
\begin{align}
    \Psi_{00} &= \phi_0(r_1)\phi_0(r_2) \otimes S(s_1,s_2),\\ S(s_1,s_2) &= \frac{1}{\sqrt{2}}\left(\ket{\uparrow \downarrow} - \ket{\downarrow\uparrow}  \right) 
\end{align}
and similarly for $\Psi_{11}$. The total wavefunction can then be written as 
\begin{align}
    \Psi = c_0 \Psi_{00} + c_1 \Psi_{11}
\end{align}
with the energy expectation value
\begin{align}
    E =& \bra{\Psi}H\ket{\Psi}\\ =& c_0^2 \bra{\Psi_{00}}H\ket{\Psi_{00}} + 2c_0c_1 \bra{\Psi_{00}}H\ket{\Psi_{11}}\nonumber\\ &+ c_1^2 \bra{\Psi_{11}}H\ket{\Psi_{11}}\nonumber.
\end{align}
The initial orbitals obtained with an MP2-PNO approach essentially assume $c_0>>c_1$ -- the total wavefunction is dominated by a single configuration $\Psi_{00}$ whereas $\Psi_{11}$ is just a small perturbation. The $\phi_0$ is then determined through minimization of $\bra{\Psi_{00}}H\ket{\Psi_{00}}$ alone (Hartree-Fock) while $\phi_1$ is determined by keeping only first order terms (in the two body interaction) in $c_1$, hence disregarding most of $c_1^2 \bra{\Psi_{11}}H\ket{\Psi_{11}}$. If the underlying assumption $c_0>>c_1$ holds this is physically justified, in the dissociated case we have however $c_0 = c_1$ leading to an unbalanced treatment in the determination of both orbitals -- as a result of neglecting the coupling between $\phi_1$ and $\phi_0$ in the determination of $\Psi_{00}$, and of neglecting the coupling between $\phi_1$ and $\phi_0$ in the determination of $\Psi_{11}$. \\

\begin{figure*}
\centering
\includegraphics[width=0.325\textwidth]{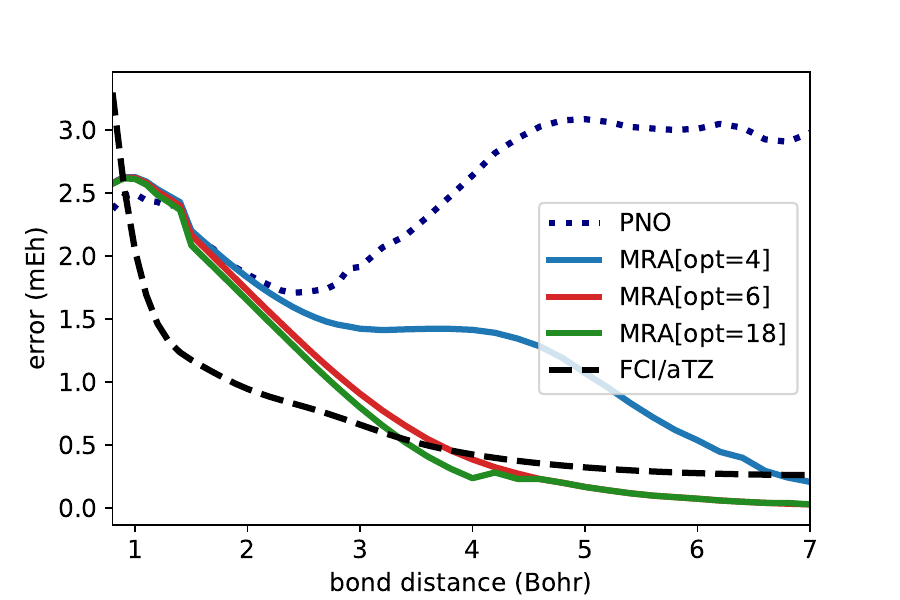}
\includegraphics[width=0.325\textwidth]{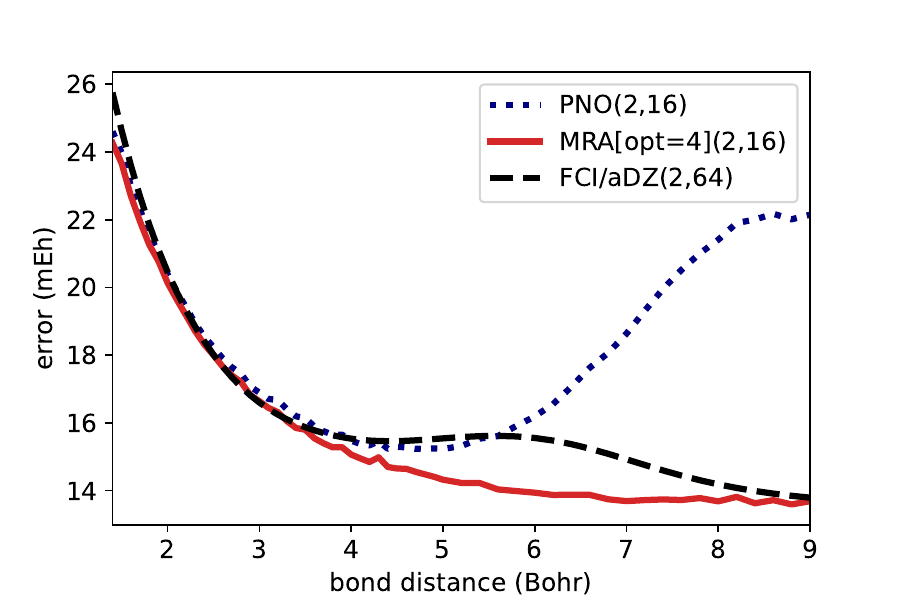}
\includegraphics[width=0.325\textwidth]{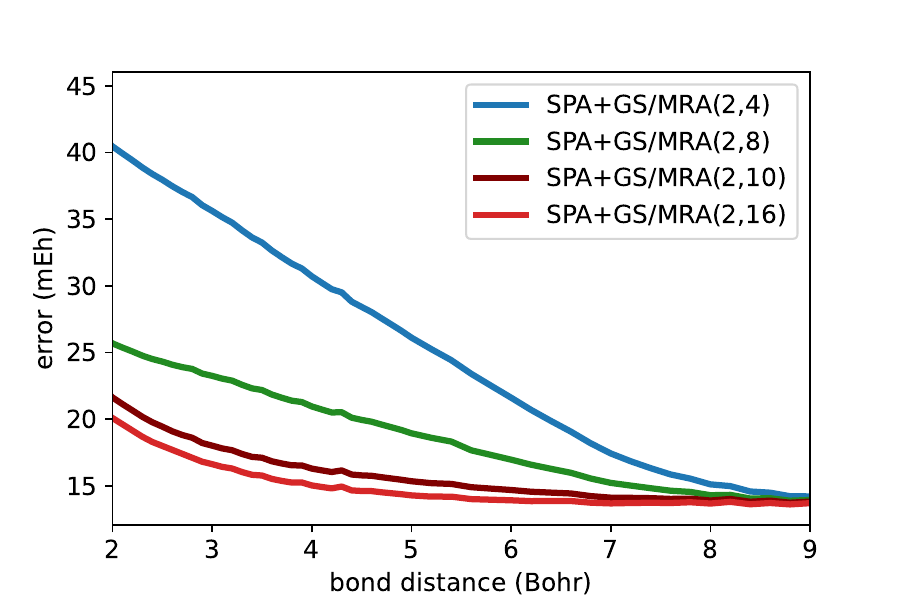}
\captionof{figure}{\textbf{Effect of selective refinement:} Orbitals of an initial orbital set (in the form of PNOs) are gradually refined while the others remain in their initial form (up to orthogonalization effects). Left: Absolute errors with respect to FCI/aug-cc-pV5Z for H$_2$, Center: Absolute errors with respect to FCI/aug-cc-pVQZ for LiH, Right: Absolute errors of LiH (same reference values) but here the number of fully-refined MRA orbitals is gradually increased. The red line, SPA+GS/MRA(2,18), is identical to the plot in the center. For H$_2$ (left) the difference of the reference values to the next basis set (aug-cc-pV5Z) is consistently below the millihartree threshold. For LiH aug-cc-pVTZ was chosen as a reference as it contains a similar amount of orbitals on Li as aug-cc-pVQZ on H.}\label{fig:lih-summary}
\end{figure*}

In Fig.~\ref{fig:lih-summary} a similar behavior can be observed for LiH. Here, the total errors are larger as we are dealing with a more correlated system, although the dissociated LiH still contains only two active electrons. In LiH, we assumed a frozen core approximation, keeping the two core (1$s$) electrons of the Li atom frozen in all calculations. For a discussion on the validity of this approximation even for orbital bases that allow core-correlation see the appendix of ~\cite{kottmann2022molecular}. 

As an alternative to enlarging the active space and optimizing all orbitals, only a limited number of active orbitals can be optimized as the active space increases. This possibility was also investigated using a (2, 18) active space for H$_2$ and a (2, 16) active space for LiH (see Fig.~\ref{fig:lih-summary}). Here, only a fixed number of the initial NOs (ordered by their respective occupation numbers) are refined. 

The examples illustrate that it is not always necessary to refine all initial orbitals and thus the computational cost of the refinement can potentially be reduced. In our current implementation, a limitation to a fixed number of orbitals does however not result in considerable savings in computational time. Furthermore, orbitals that are not optimized are also modified in each optimization cycle to ensure orthonormalization, which entails that the one- and two-electron integrals of these orbitals must also be updated in each cycle nonetheless -- and this step can become computationally quite expensive.

\subsection{BeH$_2$}

\begin{figure*}
\centering
\includegraphics[width=0.325\textwidth]{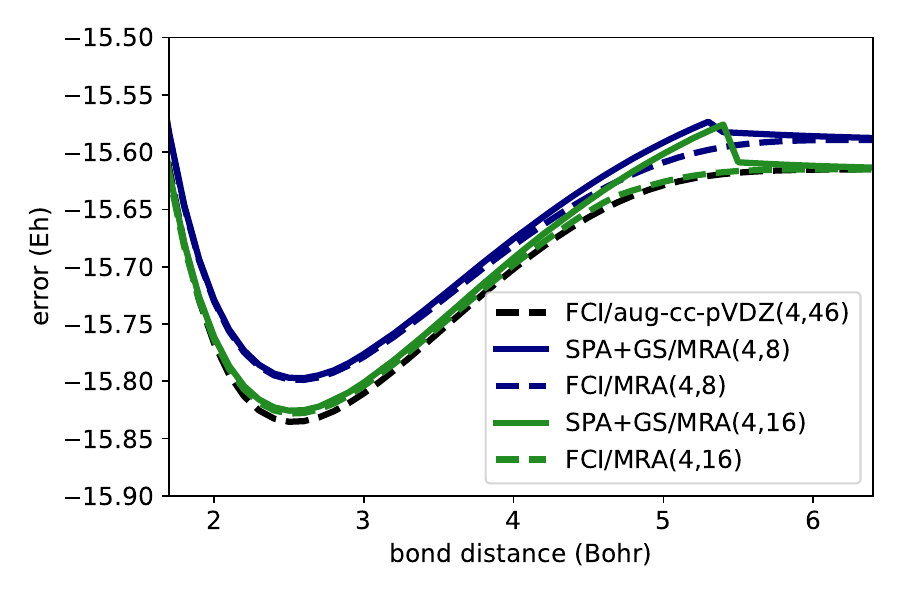}
\includegraphics[width=0.325\textwidth]{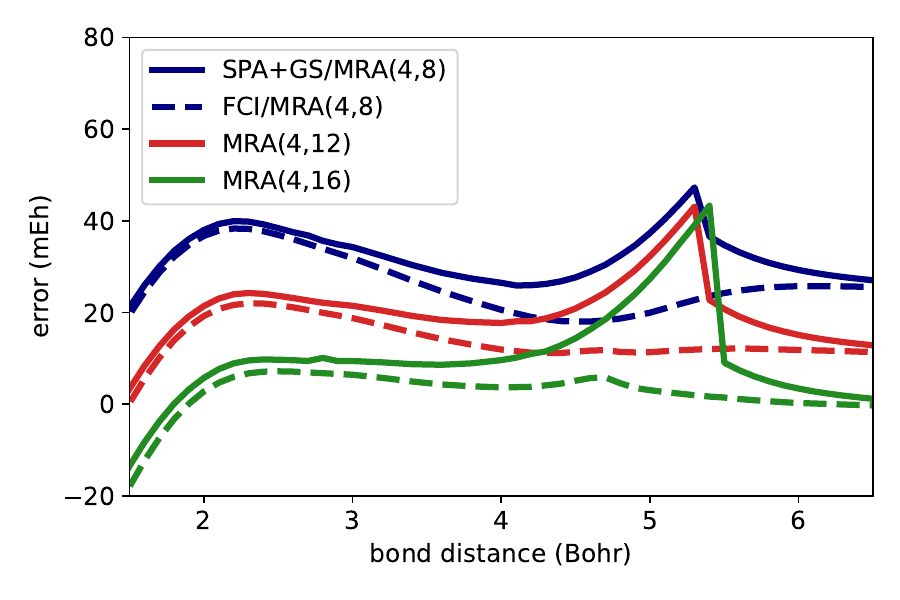}
\includegraphics[width=0.325\textwidth]{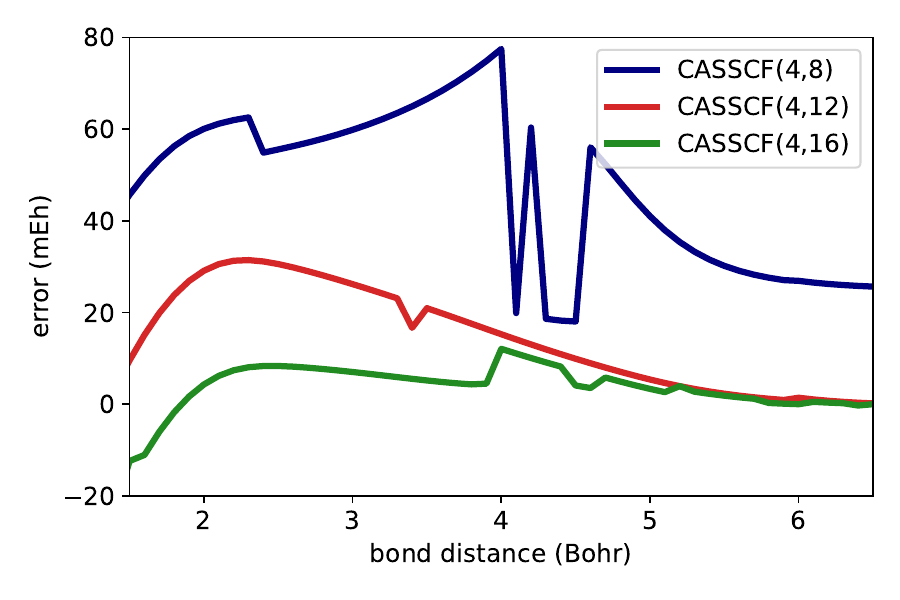}
\captionof{figure}{
BeH$_2$ in MRA orbitals determined by VQEs (SPA+GS). Left: Absolute energies. Center: Energy errors relative to FCI/cc-pVDZ(4,46). Right: CASSCF/aug-cc-pVTZ with different sizes labeled as ($N_e$, $2N_o$) using the number of spin-orbitals $2N_o$ for consistency in notation.}\label{fig:BeH$_2$}
\end{figure*}

As a further example, we studied the dissociation of BeH$_2$ by means of SPA+GSD/MRA and CASSCF calculations employing active spaces of different sizes. In contrast to H$_2$ and LiH, BeH$_2$ has four valence electrons, sucht that the SPA+GSD approach is no longer exact by construction which allows us to study the effects of an imperfect ansatz on the refinement of the orbital basis. The dissociation curves with all three methods for the active spaces (4, 8), (4,12) and (4,16) are summarized in Fig.~\ref{fig:BeH$_2$}.

A comparison of SPA+GSD/MRA with FCI/MRA shows that the SPA+GSD approach can reproduce the FCI/MRA energy almost exactly for all active spaces, both around the equilibrium distance and for high distances in a dissociated situation. This can be explained on the basis of valence-bond resonance structures where the ``molecular'' one dominates around the equilibrium while the ``atomic'' resonance structure dominates at the dissociated limit. An SPA ansatz can represent only a single resonance structure, so its associated error will decrease within the dissociation curve. For a further in-depth discussion, we refer the interested reader to Ref.~\cite{kottmann2024quantum}, in particular Fig.~5 and its accompanying text therein.\\

Turning next to a comparison of FCI/MRA with CASSCF/aug-cc-pVTZ shows a consistent behavior of the refined bases with respect to its traditional counterpart. In passing we note a challenging convergence of the smallest active space for the CASSCF with default parameters (\textsc{pyscf} implementation) which can be attributed to a convergence to local minima often encountered for large and diffuse Gaussian basis sets like the one we employ in the current case (aug-cc-pVTZ). Convergence to local minima could then be avoided if one employs active-space selection tools like the ones detailed, for example, in Ref.~\ \citenum{stein2016automated}\ and \citenum{sayf17a}, respectively. 

\subsection{H$_4$}\label{sec:h4}

\begin{figure*}
\centering
\begin{tabular}{cc}
\includegraphics[width=0.45\textwidth]{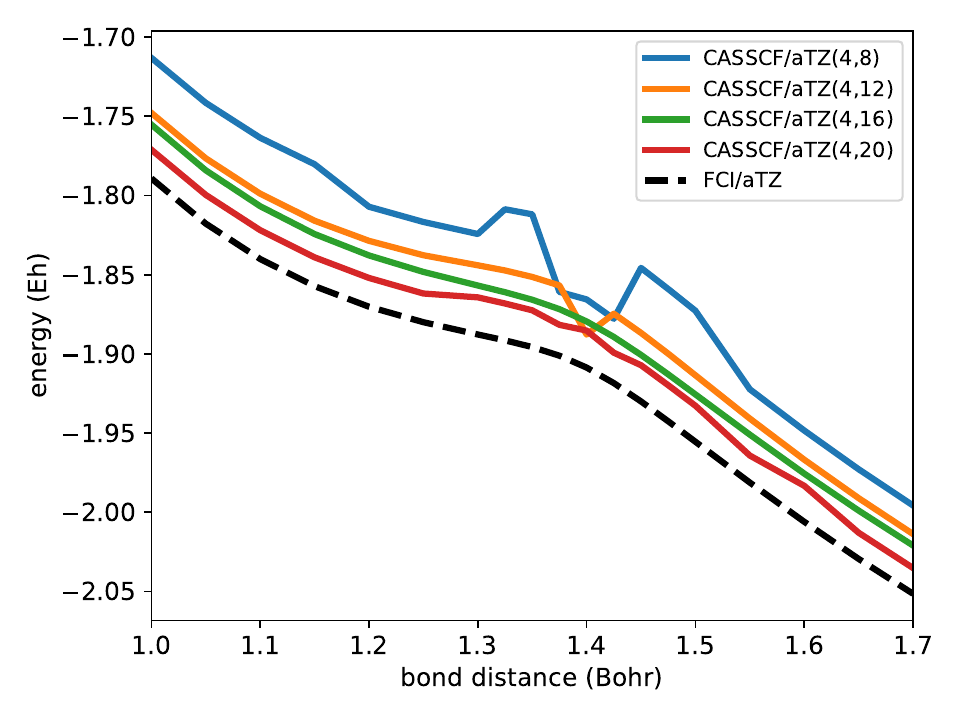}&
\includegraphics[width=0.45\textwidth]{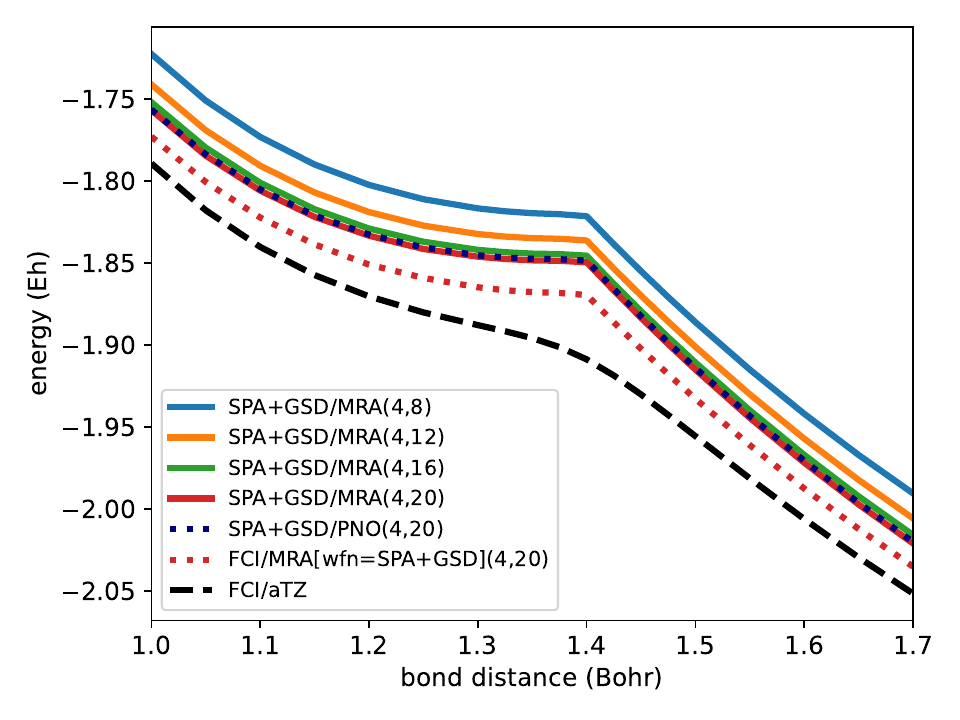}\\
\includegraphics[width=0.45\textwidth]{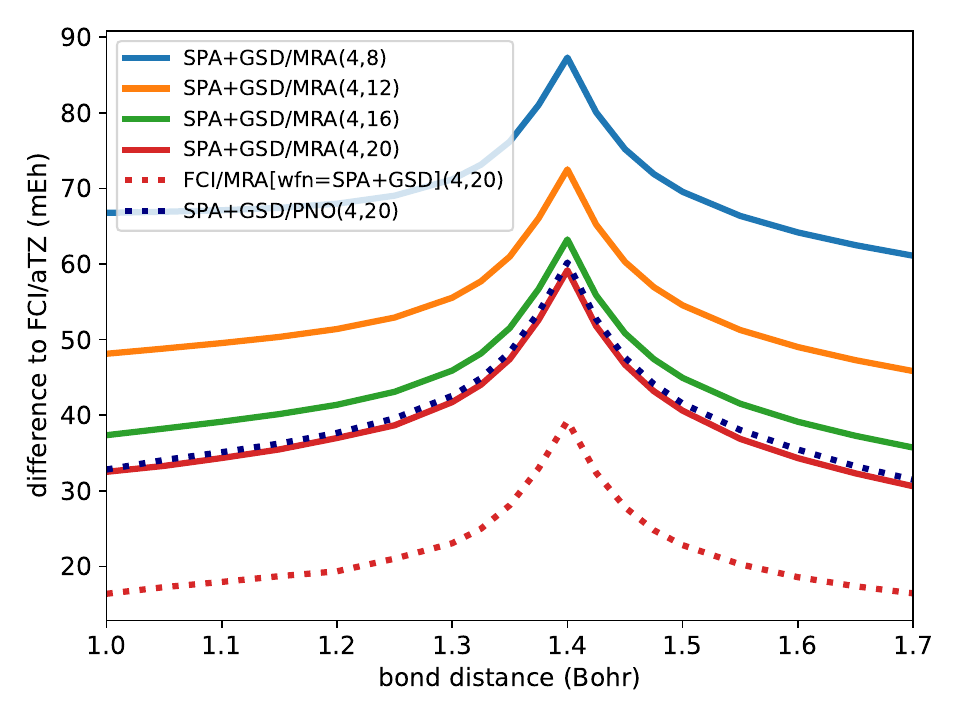}&
\includegraphics[width=0.45\textwidth]{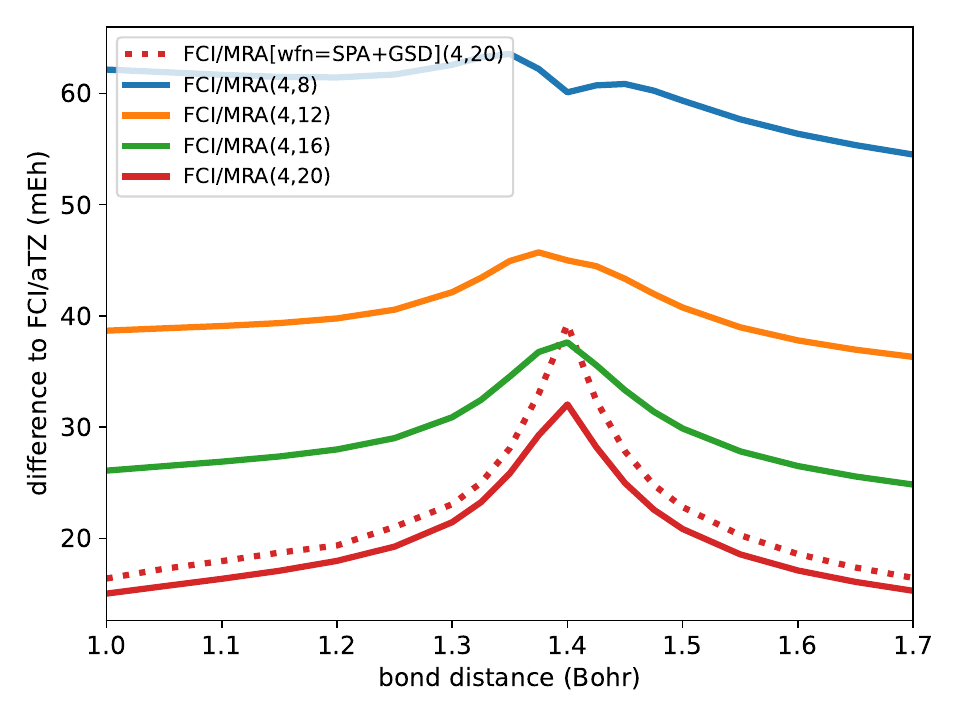}\\
\end{tabular}
\captionof{figure}{\textbf{Comparison of methods for H$_4$:} Comparison between CASSCF/aug-ccpvtz(a) with SPA+GSD(b) and FCI(c) each with optimized MRA orbitals for the dissociation of two parallel hydrogen molecules for different active spaces. All calculations use PNOs (see Fig.~\ref{fig:H4_2}) as initial guesses.}\label{fig:H4_1}
\end{figure*}

As a final example, we will study the potential energy surface for the Paldus~\cite{paldus1993h4} H$_4$/(H$_2$)$_2$ system. Similar to BeH$_2$ the SPA ansatz is unable to describe the full potential energy surface with the same accuracy, leading to a cusp-like feature in the resulting energy error around the square structure. The underlying reason comes from the inability of the SPA ansatz to describe more than a single valence-bond resonance structure while the square configuration has two degenerate resonance structures.~\cite{kottmann2024quantum}
The system here was specifically selected because of this shortcoming of the quantum circuit.
SPA+GSD/MRA, FCI/MRA and CASSCF calculations were performed with the active spaces (4,8), (4,12), (4,16) and (4,20) 
with the potential energy curves around the square configuration and the results are summarized in Fig. ~\ref{fig:H4_1}. Regardless of the method, a significant reduction in the energy over the entire potential energy curve can be seen with increasing active spaces. Notably, the energy does not yet appear to converge with the active space size even for the largest active spaces. As expected, the deviation in energy is particularly large around the square configuration, as more orbitals are required to correctly describe the increasingly complex chemical bonding situation.

Overall, MRA methods provide continuous energy profiles and almost consistently lower energies than the CASSCF counterpart, where significant energy jumps occur in some cases highlighting difficulties in converging the CAS wavefunction  without manual (or automated) adjustment of the CAS space orbitals. A comparison between SPA+GSD calculations with non-optimized orbitals (SPA+GSD/PNO) and optimized orbitals (SPA+GSD/MRA) shows hardly any differences at the example of the (4,20) active space, indicating that the HF orbitals and PNOs were already almost optimal for the problem at hand. If the SPA+GSD solver is  replaced by FCI based on the previously optimized SPA+GSD/MRA orbitals (FCI/MRA[wfn=SPA+GSD]), lower energies are obtained, although the kink in the quadratic configuration remains albeit slightly mitigated. This finding indicates that the shortcomings of the SPA ansatz were inherited with  the refined basis in a more drastic way, as for example with the BeH$_2$ molecule.

\begin{figure*}
\centering
\includegraphics[width=0.9\textwidth]{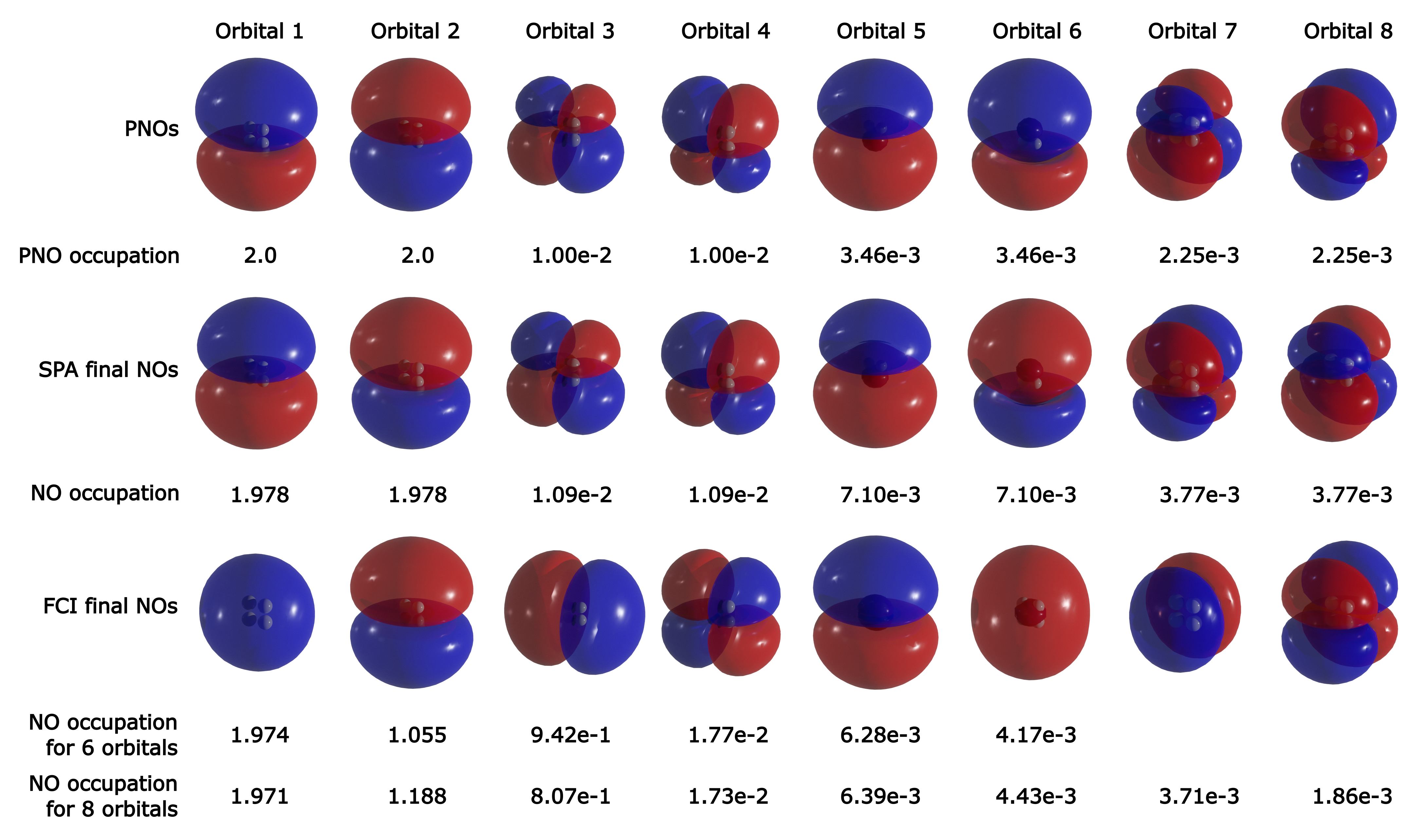}
\captionof{figure}{\textbf{MRA orbitals for H$_4$:} Initial HF orbitals and PNOs as well as the optimized natural orbitals and their occupation resulting from a SPA+GSD or FCI calculation with 2 active electrons in 8 (6) spatial orbitals for a quadratic H$_4$ configuration.}\label{fig:H4_2}
\end{figure*}

When switching to an FCI based refinement, there is still an observable deficiency around the square configuration where the energetic lowering with increasing basis set size seems to be hindered. 
The visualization of the active orbitals after orbital optimization provides an insight into the reason for this behavior. For H$_4$ in the square configuration with a (4, 16) active space, all active spatial orbitals before optimization including their PNO occupation as well as the NOs optimized with SPA+GSD/MRA and FCI/MRA including occupation numbers are depicted in Fig.~\ref{fig:H4_2}. 

With the exception of phase changes and the exchange of the order of equally occupied orbitals, there are no significant differences between the original PNOs and the final NOs from the MRA-SPA calculation. Although the orbitals were refined and the energy lowered, SPA is not able to describe the binding situation correctly. In SPA, the system is treated as two parallel H$_2$ molecules with one bond each and thus only one of the two possible resonance structures is described accurately. This limitation is particularly evident in the NO populations. Thus, one set of four orbitals is obtained for each H$_2$ bond with the NOs having the same populations for both bonds. By contrast, FCI-MRA, can describe both resonance structures simultaneously and thus the many-body wavefunction correctly. As a result, the orbital shape also changes fundamentally in the refinement process and the populations of orbitals 2 and 3 approach each other (last row in Fig.~\ref{fig:H4_2}). Ideally, both orbitals should have the same population after the refinement, but in the case of a six orbital calculation, the energy converges before both orbitals have exactly the same shape and population. In the case of a calculation with 8 orbitals, the two additionally generated PNOs favor one of the two resonance structures, which eliminates the expected degeneracy of the ground state and the populations of orbitals 2 and 3 do not equalize any further. This physically inaccurate description cannot be compensated by the orbital refinement, as it is not able to break certain symmetries. The selection of suitable initial orbitals via PNOs therefore represents a limitation in the method presented here and requires particular attention in challenging cases such as H$_4$. Interpretive concepts based on valence-bond resonance structures~\cite{kottmann2022molecular, kottmann2024quantum} provide a valuable tool in estimating these effects.

\section{Conclusion \& Outlook}
In this work, we have presented a novel approach for quantum chemistry in the framework of quantum computing that combines variational quantum eigensolvers with a basis-free representation of orbitals. Starting from initial MRA representations of Hartree-Fock orbitals and PNOs, the chosen VQE ansatz is alternately re-optimized and a refinement of the orbitals is performed until both parts converge. The resulting orbitals, dependent on the initial guess, thus obtained provide the best possible energies, while at the same time \textit{eliminating the need} to choose a basis set. Furthermore, the use of PNOs provides a systematic basis to define active spaces and avoids a time-consuming selection of a suitable and system-dependent space. In the numerical examples of this work they provide well-suited initial guesses for the full refinement on a broad range of molecules where we highlighted and analyzed exceptions and potential shortcomings of PNO guesses. The capability to calculate the best possible energy from a given number of orbitals without including unnecessary orbitals in the active space allows in most cases a reduction of the active space size without significant loss of accuracy and thus reduces the number of required qubits and gates in the case of VQEs.  
All these advantages but also current limitations were demonstrated by combining orbital refinement with the SPA+GSD VQE ansatz using various electron correlation and chemical bonding challenges at the example of a number of textbook molecular systems for quantum-chemistry developments. In most cases our VQE approach with refined orbitals led to consistently better and, above all, continuous energies over the entire dissociation profiles compared to (unguided) CASSCF reference calculations in large basis sets.
From a quantum computing point of view, all examples discussed in this work exemplify how  results obtained with a VQE ansatz based on MRA-PNOs can be improved without modifying the ansatz itself, making the ansatz with its low circuit depth a promising candidate for meaningful quantum chemistry calculations on real, contemporary quantum hardware.

Even though the code for orbital refinement developed in this work stands on the shoulders of the high-performance library \textsc{MADNESS}\cite{harrison2016madness}, it is not yet fully optimized, which means that the number of possible orbitals is lower than it could be. Nevertheless, the code is suitable for making use of significantly larger active spaces than those demonstrated in this work. Work along these lines is currently in progress and will be published elsewhere. Notably, the initial MRA-PNO code can, for example, already be routinely employed to compute systems with hundreds of orbitals~\cite{kottmann2020direct}. In the foreseeable future, the limiting factor is therefore the solution of the many-particle wavefunction and the simulation/execution of/on a quantum computer.
Furthermore, we would like to point out that the combination of orbital refinement with an initial active space comprised of PNOs as demonstrated in this work as well as the newly developed code base is not limited to the use of VQEs but can also effortlessly be combined with other classical active space solvers.~\cite{valeev2023direct}

\section*{Acknowledgement}
JSK gratefully acknowledges funding from the Hightech Agenda of the state of Bavaria and the resources on the LiCCA HPC cluster of the University of Augsburg, co-funded by the Deutsche Forschungsgemeinschaft (DFG, German Research Foundation) – Project-ID 499211671. The idea for this work was born at the ``Numerical Methods in Quantum Chemistry" workshop 2023 in Troms{\o}, Norway. SK \& JSK would therefore like to thank the organizers for a truly inspiring conference in an astonishing setting beyond the northern arctic circle. SK and FL furthermore acknowledge fruitful discussions with Roberto di Remigio Eik{\aa}s and Leander Thiessen.  

\bibliography{main}

\clearpage

\appendix
\section{Computational Details and Notation}\label{sec:comp-details}
In the following, we provide further examples to clarify the used notation of the methodology:
FCI/MRA(4,8)[opt=4,it=2]-SPA+GSD denotes a FCI calculation with $4$ active electrons within an MRA orbital basis that was refined using a SPA+GSD wavefunction and where the $4$ of the $8$ original spin orbitals (here always MRA-PNOs) where refined -- here the refinement is stopped after 2 macro-iterations. We stay consistent with previous literature and count in spin orbitals -- the refinement is however defining the spatial orbitals only. For brevity we just write MRA if the wavefunction used for orbital refinement is identical to the wavefunction used in the final energy calculation: E.g. We write FCI/MRA(4,8) instead of FCI/MRA(4,8)-FCI.

To study different properties of the proposed refinement approach such as the accuracy or the convergence behavior with respect to the number of iteration steps or the size of the active space, a number of small molecular systems were employed. Regardless of the molecule, all MRA calculations were performed in a cubic simulation box [-50 a.u., 50 a.u.]$^3$ with a polynomial order of the wavelets k=7 and a accuracy threshold of 1e-4. The initial MRA-HF orbitals, MRA-PNOs and corresponding molecular orbitals were generated via the Tequila~\cite{tequila} interface, which utilizes \textsc{MADNESS}~\cite{harrison2016madness} for this purpose. The MRA-PNOs are optimized according to Ref.~\cite{kottmann2020reducing}. Tequila is also used for the creation and execution of the VQE circuits, whereby the UpCCGSD~\cite{lee2018generalized} in the form of Ref.~\cite{kottmann2022optimized} model was used. For the Fermion to qubit mapping, Tequila relies on Openfermion~\cite{OpenFermion} and on Qulacs~\cite{qulacs} as quantum simulator backend. The initial VQE parameters are randomly generated 10 times each for H$_2$ and LiH and 20 times each for BeH$_2$ and H$_4$ and then optimized. The parameter set that delivers the lowest energy is used in the subsequent steps.

In the orbital optimizer code, \textsc{MADNESS} is used for all operations related to MRA orbitals and the eigen library is used for all linear algebra operations. Input files are read in JSON format using the JSON for Modern C++~\cite{Lohmann2022} library. In the examples shown, all valence HF orbitals and PNOs were rotated into the base of the NOs, unless otherwise specified, and all NOs with an occupation above 1e-3 were optimized simultaneously until the norm of the deviation of all orbitals drops below a threshold of 1e-3. The one and two electron integrals are written to disk in their original orbital basis in numpy format in order to optimize the parameters of the VQE approach to the modified Hamiltonian in the next iteration step. 

Since the selected VQE ansatz is not exact for BeH$_2$ and H$_4$, the VQE was exchanged with the FCI solver from the PySCF~\cite{pyscf1,pyscf2,pyscf3} package in reference calculations. This allows to assign occurring effects to the use of optimized MRA orbitals or the VQE approach. Furthermore, complete active space self-consistent field (CASSCF) calculations were performed, which are similar to the refinement approach used in the sense that both orbitals and the many-body wavefunction are optimized simultaneously or alternately. In contrast to the presented basis set-free approach, a standard basis set is used in CASSCF. To minimize the resulting basis set error and to be able to compare the results, the aug-ccpvtz basis set was used in this work. The CASSCF calculations were also carried out with PySCF, whereby all parameters not explicitly mentioned were left at their default values. For H$_2$ and LiH, additional FCI calculations were performed using the same aug-ccpvtz basis set.

\begin{figure*}
\centering
\includegraphics[width=0.9\textwidth]{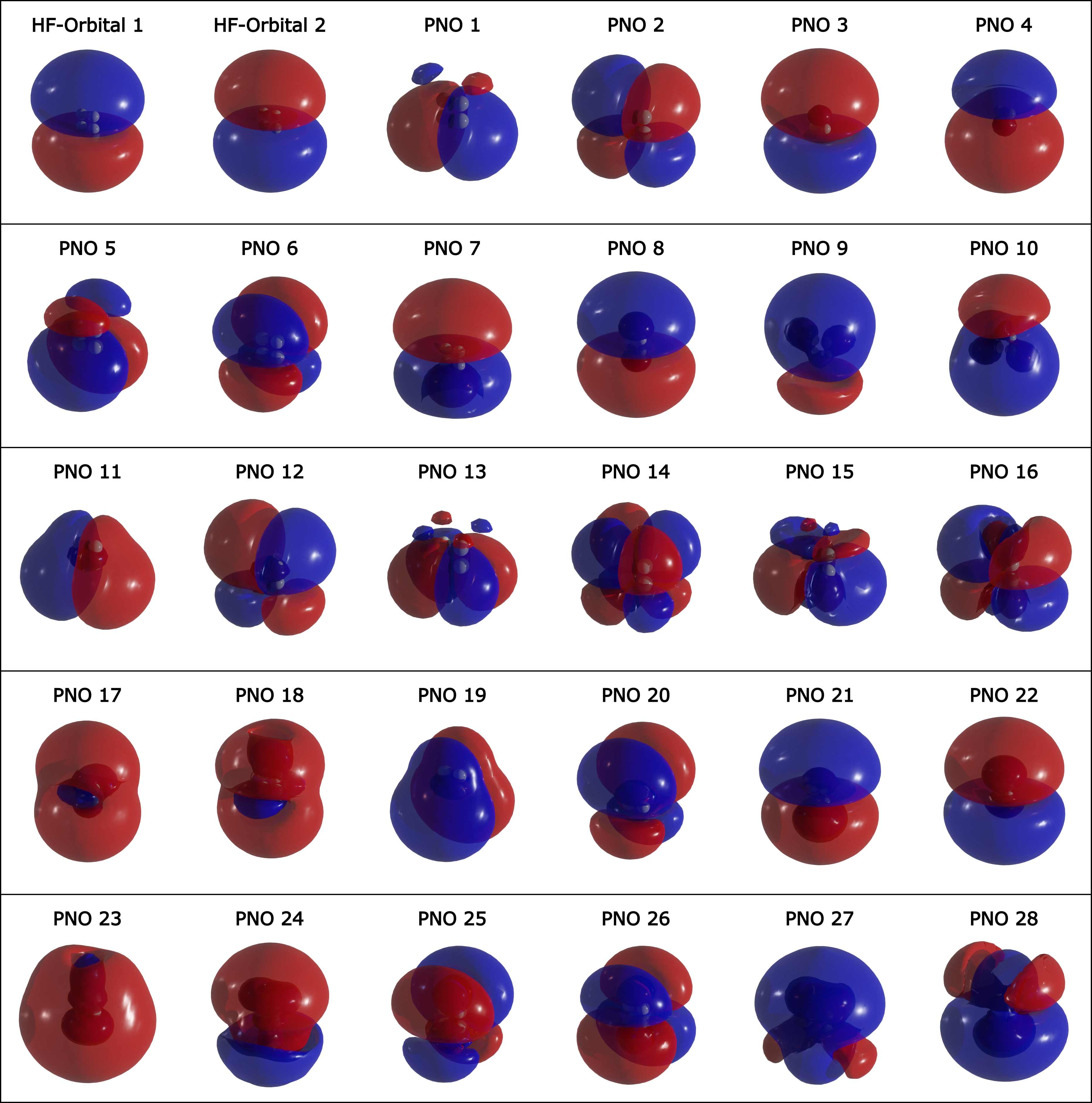}
\captionof{figure}{Hartree Fock orbitals and the first 28 pair-natural orbitals for a square H$_4$ configuration. Unlike the calculations shown in section~\ref{sec:h4}, where the orbitals were orthonormalized symmetrically, the Cholesky orthonormalization was used here.}
\end{figure*}

\end{document}